\documentclass[runningheads]{llncs}
\usepackage{graphicx}
\usepackage{rotating}
\usepackage{amsmath}
\usepackage{makecell}
\usepackage{courier}
\begin{document}
\title{Towards Tumour Graph Learning for Survival Prediction in Head \& Neck Cancer Patients}
\titlerunning{Towards Tumour Graph Learning for Survival Prediction in H\&N Cancer}
\author{Ángel Víctor Juanco-Müller \inst{1,2}\orcidID{0000-0003-2724-7903
	} \and
	João F. C. Mota \inst{2}\orcidID{0000-0001-7263-8255} \and
	Keith Goatman \inst{1}\orcidID{0000-0003-1379-1860} \and Corné Hoogendoorn  \inst{1}\orcidID{0000-0002-4914-9936}}
\authorrunning{A.V. Juanco-Müller et al.}
\institute{Canon Medical Research Europe Ltd., Edinburgh, UK \email{victor.juancomuller@mre.medical.canon}  \and
	Heriot-Watt University, Edinburgh, UK}

%
\maketitle              
\begin{abstract}
With  nearly one million new cases diagnosed worldwide in 2020, head \& neck cancer is a deadly and common malignity. There are challenges to decision making and treatment of such cancer, due to lesions in multiple locations and outcome variability between patients. Therefore, automated segmentation and prognosis estimation approaches can help ensure each patient gets the most effective treatment. This paper presents a framework to perform these functions on arbitrary field of view (FoV) PET and CT registered scans, thus approaching tasks 1 and 2 of the HECKTOR 2022 challenge as team \texttt{VokCow}. The method consists of three stages: localization, segmentation and survival prediction. First, the scans with arbitrary FoV are cropped to the head and neck region and a u-shaped convolutional neural network (CNN) is trained to segment the region of interest. Then, using the obtained regions, another CNN is combined with a support vector machine classifier to obtain the semantic segmentation of the tumours, which results in an aggregated Dice score of 0.57 in task 1. Finally, survival prediction is approached with an ensemble of Weibull accelerated failure times model and deep learning methods. In addition to patient health record data, we explore whether processing graphs of image patches centred at the tumours via graph convolutions can improve the prognostic predictions. A concordance index of 0.64 was achieved in the test set, ranking 6th in the challenge leaderboard for this task.
\end{abstract}
\section{Introduction}
Tumours occurring in the oropharyngeal region are commonly referred to as head and neck (H\&N) Cancer. In 2020 they were the third most commonly diagnosed cancers worldwide \cite{Hyuna2021}. To inform the difficult decisions that oncologists often have to make, prognosis estimation has been shown to result in better treatment planning and improved patient quality of life \cite{Johnson2020}. Therefore, automatic lesion segmentation and risk score prediction algorithms have the potential to speed up clinicians workloads, enabling them to treat more patients.

The HECKTOR challenge was conceived \cite{Andrearczyk2020,Oreiller2022} to advance the task of automatic primary tumour (GTVp) segmentation and prognosis prediction. Since the first edition in 2020, the dataset has increased from 254 to 325 cases in 2021 \cite{Andrearczyk2021}, and up to a total of 883 cases in the 2022 edition \cite{Andrearczyk2023}. Other characteristics of the present release are the lack of region of interest (RoIs) and the inclusion of secondary lymph nodes (GTVn) as segmentation targets.

This paper describes a framework for tumour segmentation and prognosis prediction consisting of three stages. First, a localization model finds the neck region in the input scans (\S\ref{subsec:loc}). Then, to obtain segmentation masks for task 1, we train a u-shaped convolutional neural network (UNet) \cite{Ronneberger2015,Isensee2021} to distinguish between tumour and background, and a support vector machine (SVM) to predict the tumour type and discard false positives (\S\ref{subsec:seg}), resulting in the semantic segmentation output for task 1, which achieves an average Dice score (DSC) of 0.57 in the test set. Finally, we explore combinations of the deep multi task logistic regression (MTLR) model, featuring CNNs and graph convolutions networks, and the Weibull accelerated failure times (Weibull AFT) method to predict the prognosis metric, which is the \emph{relapse free survival} (\S\ref{subsec:prog}), resulting in a concordance index of 0.64 in the test set. 

The experimental implementation is detailed in \S\ref{sec:exp_setup}, results are presented in \S\ref{sec:results}, and a discussion of our findings is provided in \S\ref{sec:disc_concl}.

\section{Materials \& Methods}
\begin{figure}
	\centering
	\includegraphics[scale=0.380]{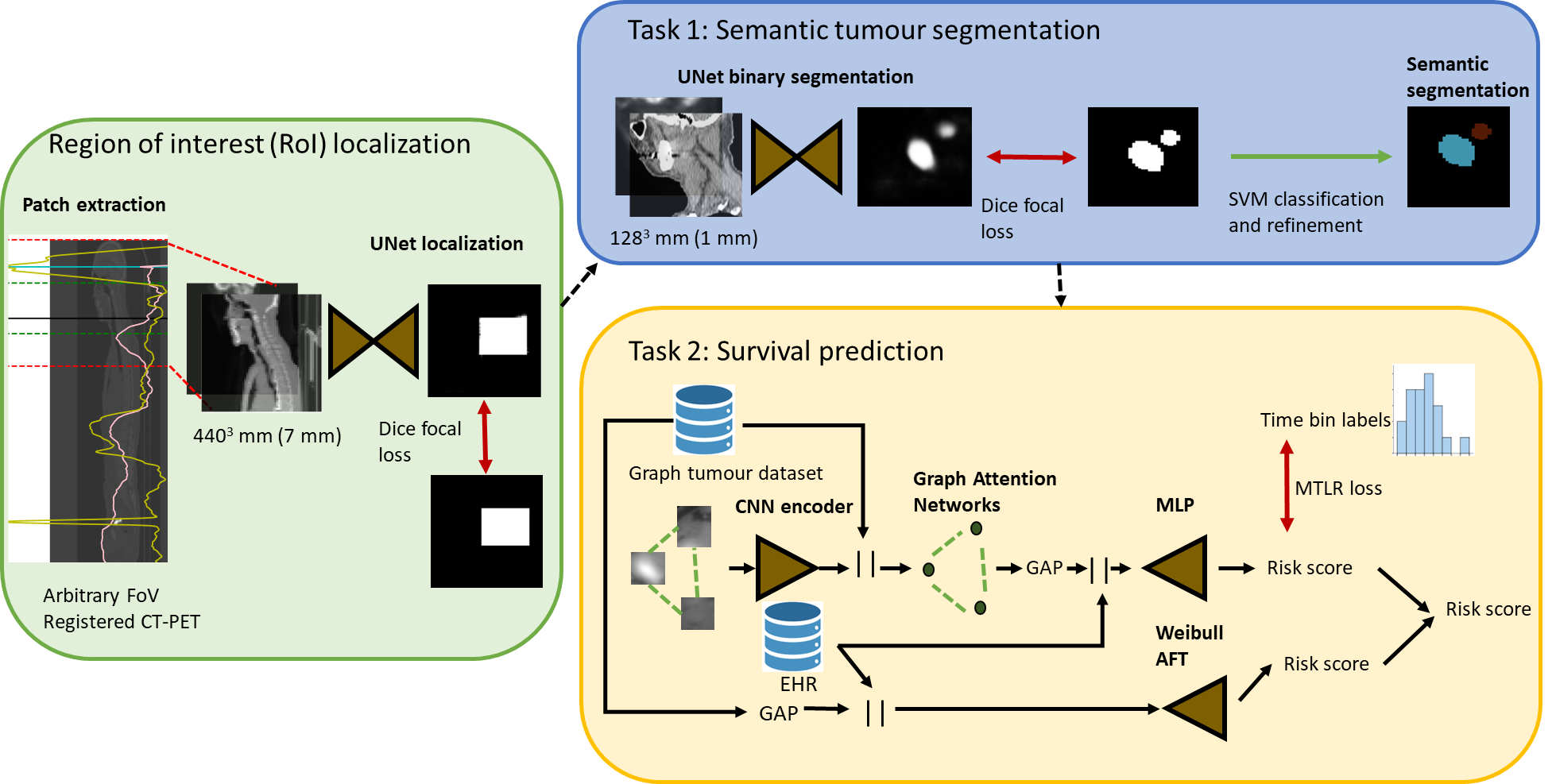}
	\caption{\label{fig:framework} The three main stages of the proposed framework. GAP stands for global average pooling, MLP for multi layer perceptron, and $||$ refers to concatenation.}
\end{figure}

This section presents the three main stages of our framework: localization \S\ref{subsec:loc}, segmentation \S\ref{subsec:seg} and survival prediction \S\ref{subsec:prog}, depicted in Fig. \ref{fig:framework}.

\subsection{Localization \label{subsec:loc}}
\begin{figure}
	\centering
	\includegraphics[scale=0.40]{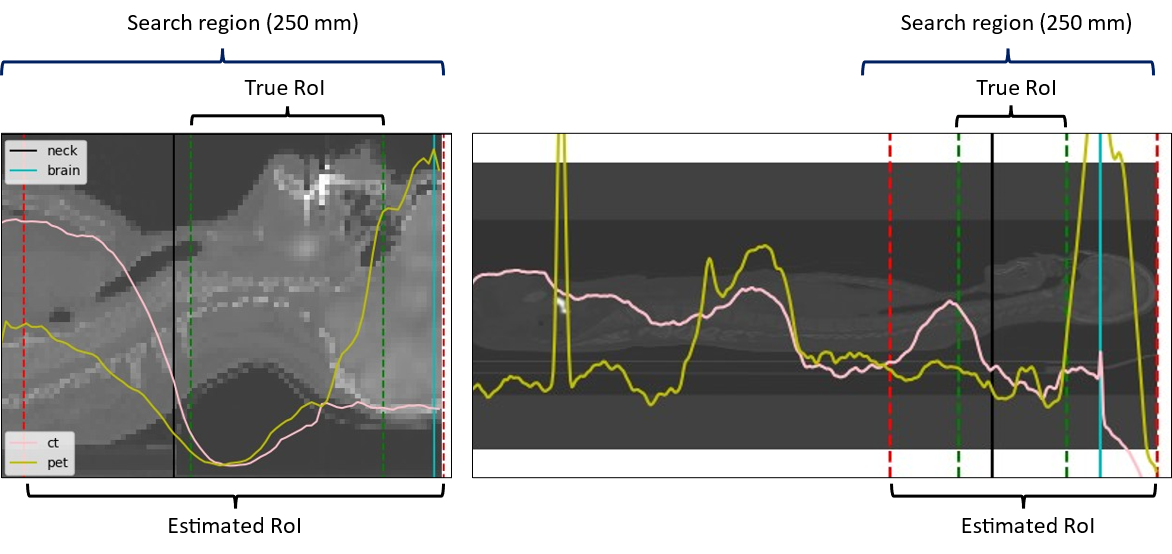}
	\caption{\label{fig:slice_range_sel} Patch extraction for two different FoV cases. The dotted red lines are the inferred bounds, whereas the green ones correspond to the ground truth location of the RoI.}
\end{figure}

First, we extract $440^3$ mm patches of the head and neck region from the arbitrary FoV PET-CT scan by analysing the CT and PET mean slice intensity along the z-axis. The brain is detected by a peak in the PET signal and the neck by an abrupt drop of the CT value. To avoid false positives caused by peaks of the PET signal in other regions of the body (e.g., bladder) we restrict the landmark search only to the first 250 mm starting from the head, as depicted in Fig. \ref{fig:slice_range_sel}.
 
\begin{figure}
	\centering
	\includegraphics[scale=0.40]{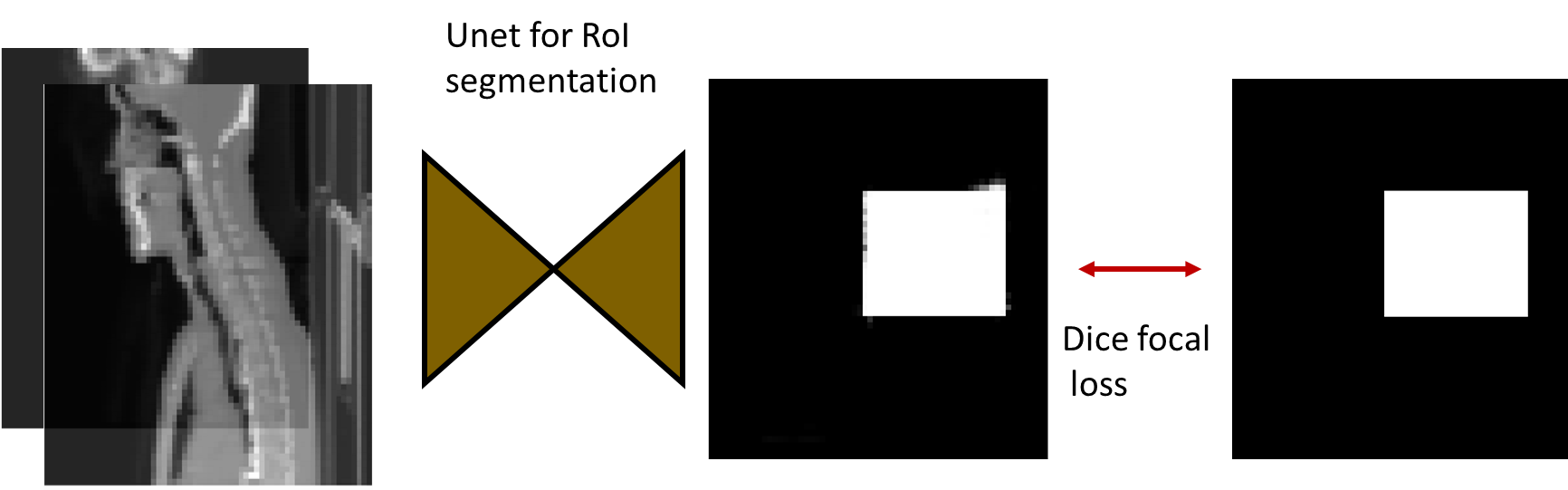}
	\caption{\label{fig:unet_roi_seg} RoI segmentation with UNet. Based on the bounds inferred in the previous step, the images are cropped to a common FoV and a UNet is trained on a low-resolution dataset to segment the target RoI bounding box.}
\end{figure}

The resulting patches are then resized to a $64^3$ mm size with trilinear interpolation for the images and nearest neighbours for the reference bounding box masks, which are obtained from the ground truth tumour segmentations. A 3D UNet \cite{Ronneberger2015,Isensee2021} is then used to segment the latter by minimizing the sum of Dice \cite{Fausto2016} and focal \cite{Lin2017} losses (Fig. \ref{fig:unet_roi_seg}), achieving a Dice score in the validation set of $0.72$. This results in a model with 3 layers, each comprising convolutional blocks, ReLU activations, and instance normalization \cite{Ulyanov2016}. From one layer to the next, we double the number of channels and reduce the spatial dimensions by half with max pooling.

\subsection{Segmentation \label{subsec:seg}}
For the segmentation task, we first apply a 3D UNet \cite{Ronneberger2015,Isensee2021} based on the model presented by the top ranked teams in 2020 \cite{Iantsen2021} and 2021 \cite{Xie2022}. The UNet has 5 levels of depth, without exceeding 320 channels, and uses residual squeeze and excitation blocks \cite{Isensee2021}. The loss function was the same one used for localization.

Because a multi-class segmentation model performed poorly, we opted instead to use the UNet for binary segmentation (tumour-background), and a traditional classifier to infer the tumour type. We tried different algorithms and a support vector machine (SVM) with radial basis functions was chosen as it yielded the best performance. The input features were: \emph{tumour centroid} and \emph{bounding box coordinates}, \emph{Euler number},	\emph{extent}, \emph{solidity}, \emph{filled area}, \emph{area of the convex hull}, \emph{area of the bounding box}, \emph{maximum Feret diameter}, \emph{equivalent diameter area}, \emph{eigenvalues of moment of inertia} and	\emph{minimum}, \emph{mean} and \emph{max values of CT and PET intensities}. 

\begin{figure}
	\centering
	\includegraphics[scale=0.52]{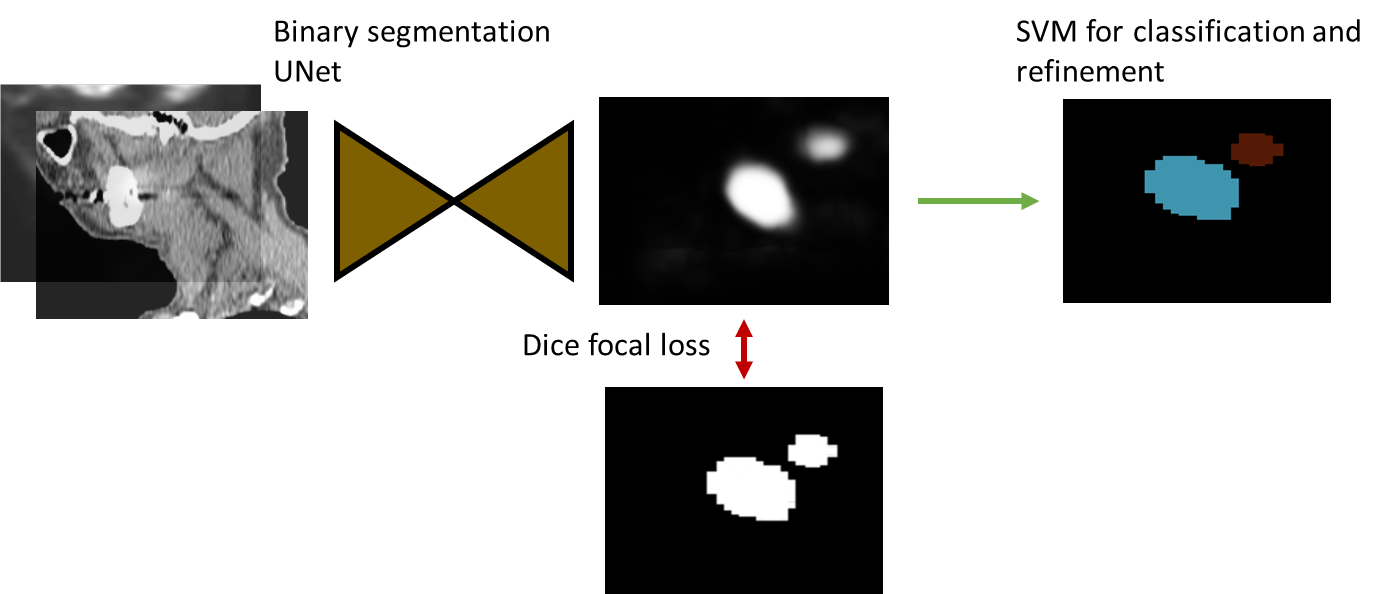}
	\caption{\label{fig:segmentation_pipeline} Semantic tumour segmentation. First a UNet segments the tumours from the background and then an SVM classifier predicts the tumour type.}
\end{figure}

The SVM classifies the tumours into three possible classes: background (to discard false positive predictions), primary tumour (GTVp) and lymph node tumours (GTVn). The input features were extracted with the Scikit-Image library \cite{VanDerWalt2014}, and the SVM implementation is provided by the Scikit-Learn \cite{Pedregosa2011} library with default parameters. The overall pipeline is depicted in Fig. \ref{fig:segmentation_pipeline}.

\subsection{Survival Prediction \label{subsec:prog}}
We first use simple models to select the most relevant tumour features for survival prediction, and then explore deep learning models that process such features together with the images and clinical data.

\textbf{Calibration Experiments}.
We considered the \emph{Cox proportional hazards} (Cox PH) \cite{Cox1972} and \emph{Weibull accelerated failure times} (Weibull AFT) \cite{Kalbfleisch1980} methods. We fitted these models on only electronic health record (EHR) data and both EHR and different combinations of the features used for tumour classification with SVM. For cases with several tumours, we considered the mean value of the features. We also included the number of tumours as an additional feature. After experimentation we identified \emph{tumour centroid}, \emph{mean CT and PET}, \emph{max CT intensities}, and \emph{number of tumours} as the combination yielding the best improvement of survival prediction in the validation set.

Among the EHR variables, there was missing information for some patients regarding alcohol and tobacco usage, performance status in the Zubrod scale, presence of human papillomavirus (HPV) and whether the patient has undergone surgery or not. Because all these variables are non-negative, we assigned the value -1 for the missing cases. This resulted in better performance than simply dropping these columns. 

The Weibull AFT model trained on EHR data and tumour descriptors achieved the highest concordance index (C-Index) in the validation set (table \ref{tab:surv_cal}). This model assumes the hazard probability to be a function of patient features $x_i$ and time $t$ parametrized by $\beta_i$ and $\rho$,
\begin{equation*}
	H(x_i,t;\beta_i,\rho) \propto -\left(\frac{t}{\textmd{exp}(\beta_0 + \sum_i \beta_i x_i)}\right)^\rho.
\end{equation*}

The log-rank test revealed that tumour descriptors like mean PET intensity and number of tumours are significant for the predictions (table \ref{tab:log_rank_test}). The parameters $\rho$ and $\beta_0$ determine the shape of the Weibull distribution and were fit to $-3.07$ and $34.24$ for the model trained only on EHR, and $-0.21$ and $10.25$ for the one trained on both EHR and tumour descriptors. We used the Lifelines package \cite{Davidson2019} implementation of Cox PH and Weibull AFT with their default parameters.
\begin{table}
	\begin{center}
		\caption{\label{tab:surv_cal} Survival calibration results in the validation set in terms of concordance index. The Cox PH and the Weibull AFT models were fitted only on EHR data and both EHR and tumour descriptors. It can be seen that Weibull AFT outperforms Cox PH and that including tumour descriptors improves the results in both cases. Best figures are in bold.}
		\begin{tabular}{|c|c|c|}
			\hline
			&  Cox PH &  Weibull AFT  \\
			\hline
			EHR & $0.66518 $ & $0.67793$ \\
			\hline
			EHR + tumour descriptors & $0.68367$ & $\textbf{0.70026}$ \\
			\hline
		\end{tabular}
	\end{center}
\end{table}
\begin{table}[t]
	\begin{center}
		\caption{\label{tab:log_rank_test} Log-rank test for the Weibull AFT model fit. The input features with lowest p-value are the most representative for the survival prediction task. Best figures in bold.}
		\begin{tabular}{|c|c|c|c|c|}
			\hline
			&\multicolumn{2}{c}{Only EHR}\vline &\multicolumn{2}{c}{EHR + tumour descriptors} \vline \\
			\cline{2-5}
			Input feature ($x_i$) &Coefficient ($\beta_i$) & p-value ($\downarrow$) &Coefficient ($\beta_i$) & p-value($\downarrow$) \\
			\hline
			Age & $-0.15$ &$0.35$ &$-0.13$& $0.23$ \\
			Alcohol & $0.36$ & $0.10$ &$0.44$& $0.07$ \\
			Chemotherapy &$-0.14$& $0.47$ &$-0.15$& $0.82$ \\
			Gender & $-0.50$ & $0.01$ &$-0.28$& $0.06$ \\
			HPV status (0=-, 1=+) & $0.37$& $0.08$ &$0.31$& $0.12$ \\
			Performance status & $-0.87$& $\mathbf{\ll 0.005}$  & $-0.69$ & $\mathbf{\ll 0.005}$ \\
			Surgery &$0.16$& $0.37$ &$0.15$& $0.65$  \\
			Tobacco &$0.13$&$0.54$ &$0.03$& $0.40$ \\
			Weight &$0.42$ & $0.03$ &$0.55$& $0.01$ \\
			Area bounding box & - & - &$-0.28$& $\mathbf{\ll 0.005}$\\
			Centroid x coordinate  & - & - &$-0.64$& $0.22$ \\
			Centroid y coordinate & - & - &$-1.41$& $0.99$ \\
			Centroid z coordinate & - & - &$0.92$& $0.23$\\
			Max CT intensity & - & - &$-0.27$& $0.05$ \\
			Mean CT intensity & - & - &$0.54$& $0.02$ \\
			Mean PET intensity & - &- &$-0.55$& $\mathbf{\ll0.005}$ \\
			Number of tumours & - & - &$-0.43$& $\mathbf{\ll0.005}$ \\
			\hline
		\end{tabular}
	\end{center}
\end{table}

\textbf{Survival Model.}
The winning method of the previous edition of the challenge, named \emph{Deep Fusion} \cite{Saeed2022}, consisted of a CNN encoder that takes a fused PET-CT image as an input, and outputs a feature embedding that is concatenated with patient EHR data. The final layer is a multi layer perceptron (MLP) connected with the multi task logistic regression (MTLR) loss, which can model individual risk scores accurately \cite{Yu2011,Jin2015,Fotso2020}. It divides the target time into bins for which survival scores are predicted, imposing constraints to deal with uncensored and censored events.

Since \emph{number of tumours}, $n$, was one of the most representative features in the calibration experiments, we hypothesize that, rather than one single image patch, processing $n$ fused PET-CT patches with graph convolution networks may provide stronger prognosis signals. In the proposed \emph{Multi-patch} model, 64-dimensional patch embeddings were first obtained with a CNN layer followed by batch normalization \cite{Ioffe2015}, ReLU activation and average pooling. Next, we built an unweighted fully connected graph of $n$ nodes, one for each of the image patches.

We assigned concatenations of the CNN embeddings of the image patches and the tumour descriptors selected in the \emph{calibration experiments} as node features, and applied two layers of graph convolution and ReLU activation to reason over the tumour graph. The improved graph attention network (GATv2) \cite{Brody2021} was chosen for its availability to perform dynamic node attention. 

\begin{figure}
	\centering
	\includegraphics[scale=0.44]{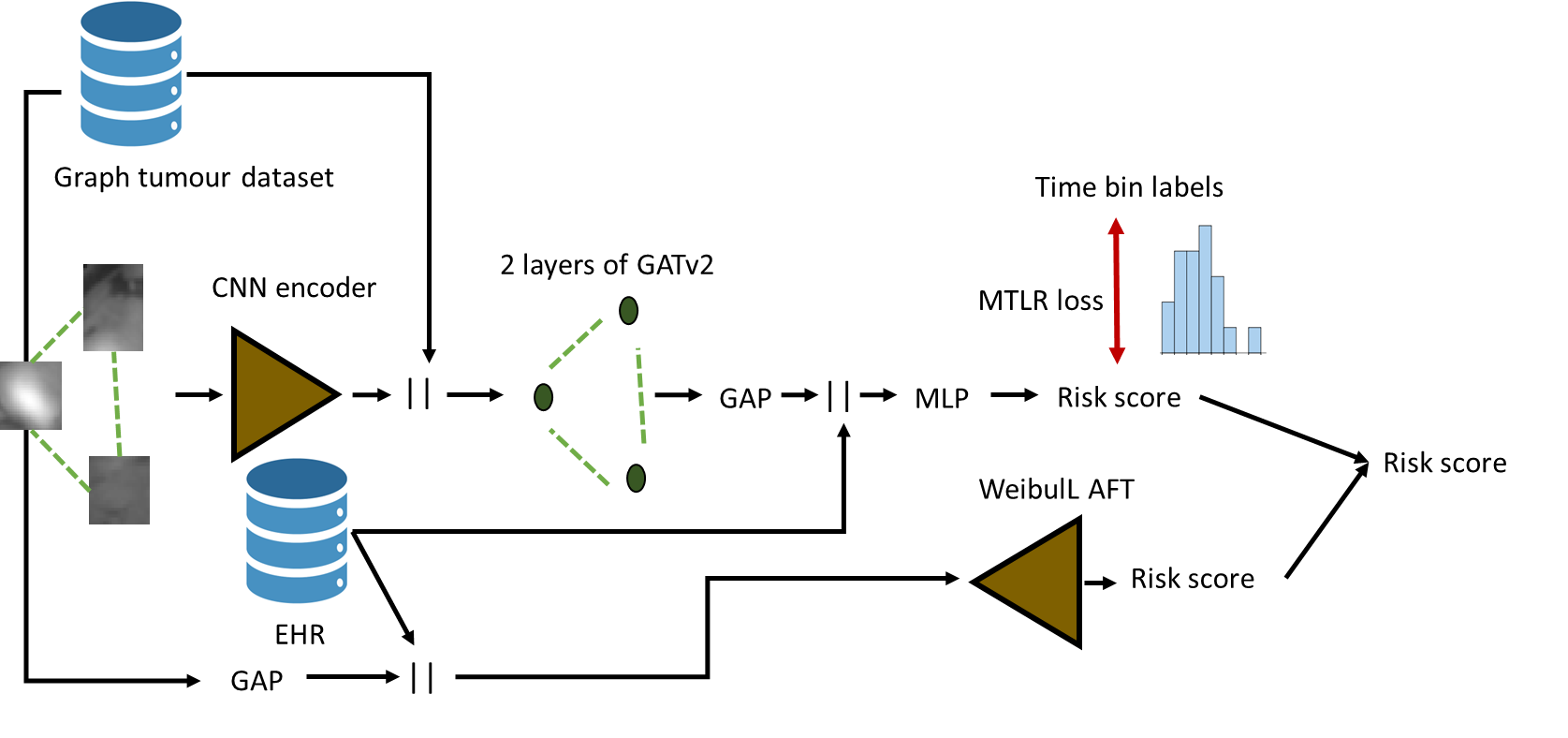}
	\caption{\label{fig:survival_pipeline} Multi-patch network for survival prediction. The inputs are fused CT-PET image patches cropped at the segmented tumour centroids, tumour features, and patient EHR data. The output is the predicted risk score.}
\end{figure}

Finally, an average pooling layer generates global graph vector embeddings, which are then concatenated with each patient's EHR data. A Multi Layer Perceptron (MLP)  produces the output logits, which are then used to compute the MTLR loss and the patient's risk score. The full pipeline is shown in Fig. \ref{fig:survival_pipeline}.

\section{Experimental Set Up \label{sec:exp_setup}}
Here we provide details of our experimental implementation, covering data splitting (\S\ref{subsec:data_spl}), the preprocessing and augmentation techniques (\S\ref{subsec:data_pr_aug}), and the hardware and network hyperparameters (\S\ref{subsec:impl_det}).

\subsection{Data Splitting \label{subsec:data_spl}}
For the segmentation task, the training data was divided into training and validation splits with 445 and 79 cases respectively for the segmentation task. To ensure a balanced representation of multi-centre data, the cases from each centre were first randomly allocated to \emph{per-centre} subsplits, and then aggregated to form the final split.

\begin{figure}
	\centering
	\includegraphics[scale=0.60]{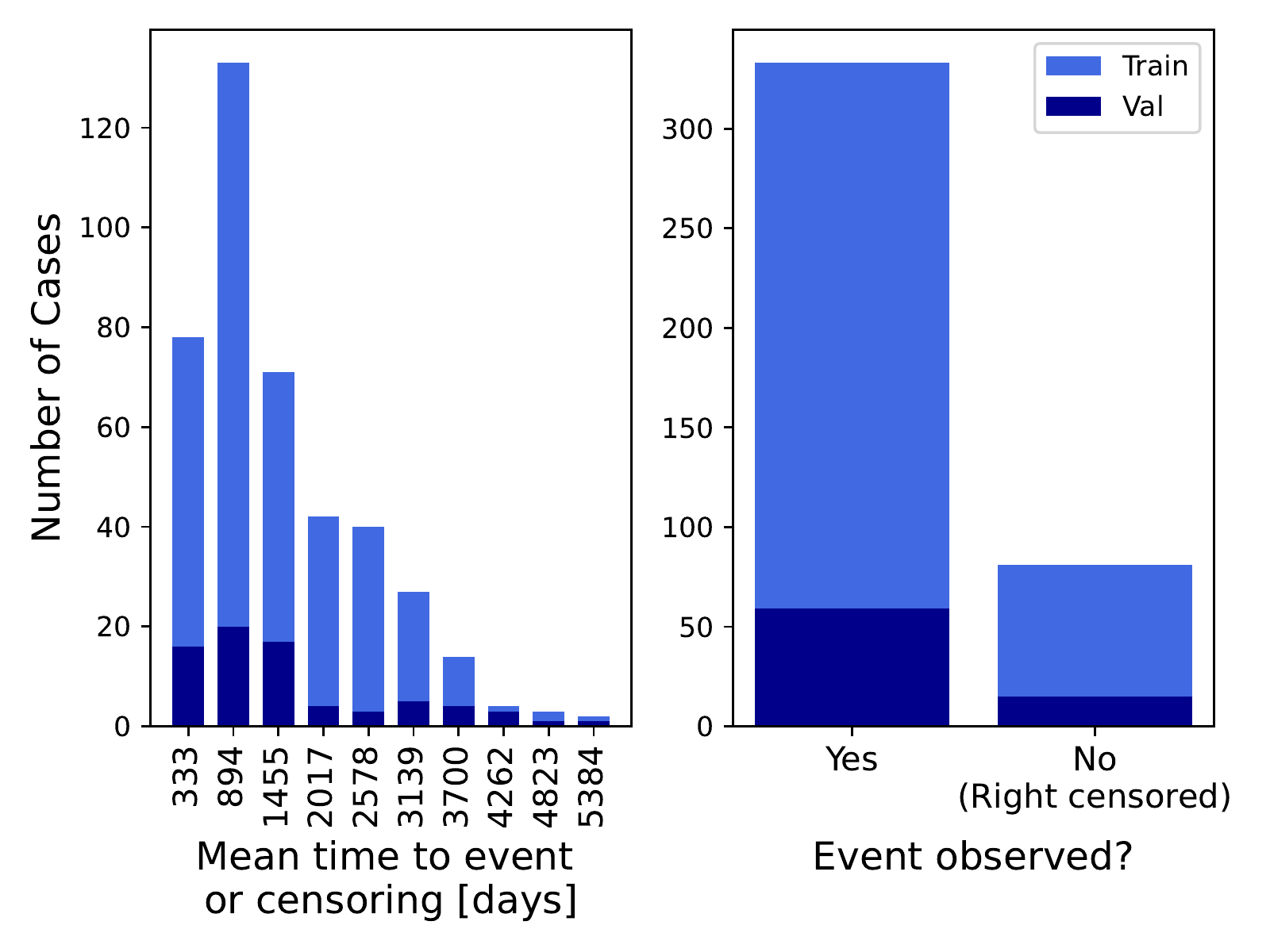}
	\caption{\label{fig:proportions_surv} Histograms of survival times and censoring status of the training and validation cases. It can be seen that both distributions are very similar, peaking at the same mean. Best seen in colour.}
\end{figure}

Since the survival dataset is a subset of the segmentation one, we opted to define the survival splits as subsets of the segmentation partitions, resulting in 414 training and 74 validation cases. In this manner the inferred risk score is determined by the segmentation inference. We confirmed (Fig. \ref{fig:proportions_surv}) the absence of important distribution shifts between the training and validation survivals times and proportion of censoring cases.

\subsection{Data Preprocessing and Augmentations \label{subsec:data_pr_aug}}
Because global context is more important than detail for localization, we resampled the images to low resolution, e.g. $(7,7,7)$ mm voxel spacing and clipped the CT values to the interval $(-1024,1024)$. Instead, for the segmentation and survival networks, where detail or texture is more important, we resampled the inputs to 1 mm isotropic voxel size and windowed the CT values to  $(-200,200)$ to enhance soft tissues. 

In all cases, the images were normalized by subtracting the mean and dividing by the standard deviation. The input of the localization and segmentation networks are pairs of CT and PET images, whereas for the survival network, the PET and CT image are fused by averaging.

For the localization model, the following random augmentations (with probability $p$) were applied the training samples: random intensity shifts in the range $(-0.5,0.5)$ ($p=0.5$), random scale shifts with in the range $(-1,1)$ for the PET and $(-0.25,0.25)$ for the CT ($p=0.5$), and random Gaussian Noise with mean $0$ and standard deviation $0.1$ ($p=0.1$). The same augmentations reported in \cite{Xie2022} were used to train the binary segmentation network. No augmentations were used for the survival prediction models.

\subsection{Implementation Details \label{subsec:impl_det}}
All models were run in a 32 GB NVidia Tesla V100, although they had different memory footprints and training times. The localization model was the lightest, whereas the binary segmentation model was the heaviest and with longer training time. All models were implemented using PyTorch \cite{Paszke2019}, PyTorch Lighting \cite{Falcon2019} and PyTorch Geometric \cite{Fey2019}. Table \ref{tab:impl_det} summarizes the different hyperparameters and other training details for these three models.
\begin{table}
	\begin{center}
		\caption{\label{tab:impl_det} Implementation and training details of the different trained networks in this study.}
		\begin{tabular}{|c|c|c|c|c|}
			\hline
			& Detection & Segmentation & \thead{Survival \\ (Deep Fusion)} & \thead{Survival\\(Multi-patch)} \\
			\hline
			Optimizer & Adam & SGD with momentum & Adam & Adam\\
			\hline
			Scheduler & \thead{Reduce LR \\ on plateau} & Poly LR & Multi step LR & Multi step LR \\
			\hline
			\thead{Initial \\ learning rate} & 0.001 & 0.001 & 0.016 & 0.016\\
			\hline
			\thead{Loss function} & Dice focal & Dice focal & MTLR &MTLR \\
			\hline
			Epoch & 100 & 100 & 100  & 100\\
			\hline
			\thead{GPU \\ RAM (GB)} & 4 & 26 & 10 & 10 \\
			\hline
			\thead{Patch size} & $64^3$& $128^3$ & $80\times80\times50$ & $32^3$ \\
			\hline
			\thead{Training \\ time (hours)} & 2 & 48 & 10 & 10\\
			\hline    
			\thead{Validation \\ metric} & \thead{Average \\ precision} & \thead{Average \\ precision}  & \thead{Concordance \\ index} & \thead{Concordance \\ index}\\
			\hline    
		\end{tabular}
	\end{center}
\end{table}
\section{Results \label{sec:results}}
Here we present our results, first for the segmentation task (\S\ref{subsec:seg_res}), then for the survival prediction task (\S\ref{subsec:surv_res}).

\subsection{Segmentation Results \label{subsec:seg_res}}
First, we assessed the binary segmentation and classification performance separately in the validation set. The binary segmentation network achieves a Dice score of $0.636$, whereas for the tumour classification problem, the macro and micro F1 scores are $0.843$ and $0.861$ respectively.

Then, we obtained the semantic segmentation outputs from the binary segmentations and classifications results, and computed the aggregated Dice score. All the intersections are divided by all the unions in the considered data split independently for each class, and then the mean of the two is computed. 

Table \ref{tab:seg} reports this metric in the validation and test sets. Although the results suggest the proposed model is not a strong segmentor, it provides suitable input for the survival task, which benefits from the features extracted from the segmented tumours.
\begin{table}
	\begin{center}
		\caption{\label{tab:seg} Results of the proposed segmentation method in the validation and test sets.}
		\begin{tabular}{|c|c|c|c|}
			\hline
			 Dataset Split &GTVp Dice &GTVn Dice   &Aggregated Dice \\
			\hline
			 Validation Set (74 cases) &  0.68514  & 0.62648 & 0.65581\\
			 \hline 
			 Test Set (359 cases) & 0.59424 &0.54988 & 0.57206\\ 
			 \hline
		\end{tabular}
	\end{center}
\end{table}

\begin{figure}
	\centering
	\includegraphics[scale=0.60]{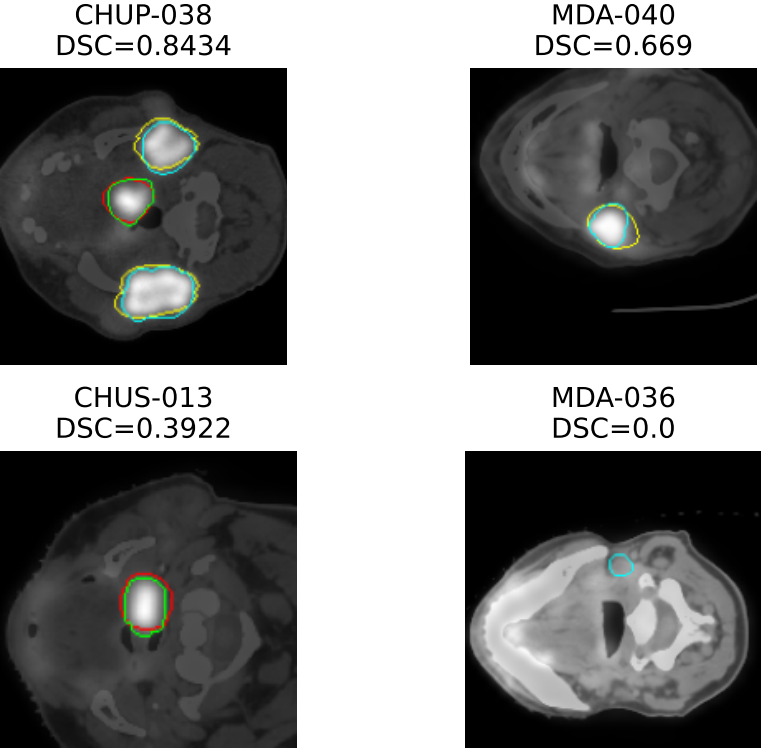}
	\caption{\label{fig:best_avg_worst} Best, two average and worst segmentation outputs. The same preprocessing and averaging technique for the survival prediction is applied to the CT and PET images. Ground truth for GTVp and GTVn contours are in green and cyan, whereas predicted contours are in red and yellow, respectively.}
\end{figure}

Finally, we qualitatively assessed the algorithm outputs by looking at the best, average and worst cases (Fig. \ref{fig:best_avg_worst}). As a result of mistakes in the localization step, some of the predicted bounding box were slightly shifted from the actual RoI, which in turn resulted errors in the segmentation stage. For example, in \emph{MDA-036} the predicted bounding box included a greater part of the brain, and the segmentation network detected a small region of brain as the tumour. Therefore, post-processing of the predicted bounding box could have improved the training stability and segmentation results.
\subsection{Survival Prediction Results \label{subsec:surv_res}}
Here we present our results for the survival prediction task. Table \ref{tab:suv_res} reports the C-Index achieved by the proposed methods in the validation and test sets. For a fair comparison, we implemented and reported the results of \emph{Deep Fusion}, the best performing neural network presented last year \cite{Saeed2022} for this task.
\begin{table}
	\begin{center}
		\caption{\label{tab:suv_res} Results of different methods in the validation set in terms of Concordance Index (C-Index). Best results in bold.}
		\begin{tabular}{|c|c|c|}
			\hline
			Model & Validation Set (74 images) &Test Set (339 images)\\
			\hline
			Weibull AFT &  0.70026  & \textbf{0.64086} \\
			\hline
			Deep Fusion \cite{Saeed2022} &  0.60587  & 0.47923 \\
			\hline
			Deep Fusion \cite{Saeed2022} + Weibull AFT &  0.72194  & 0.64081 \\
			\hline
			Multi-patch (ours) & \textbf{0.75000} & 0.39679 \\
			\hline
			Multi-patch + Weibull AFT (ours) & 0.70536 & 0.64013 \\
			\hline
		\end{tabular}
	\end{center}
\end{table}

The proposed Multi-patch method performs best in the validation set, followed by Deep Fusion. Nevertheless, both deep learning methods generalize poorly to the test set, with our method obtaining the worst metric. To try to mitigate the overfitting, we ensembled the outputs of the deep learning and the Weibull AFT methods by simple averaging. However, even under this setting, the WeibulL AFT model alone was the best performing method in the test set. Little difference was observed between this model alone and ensembles of it and the deep learning algorithms.

\section{Discussion \& Conclusion \label{sec:disc_concl}}
We presented a framework for tumour segmentation and prognosis prediction in head \& neck cancer patients, which may have a positive impact in patient management and personalized healthcare. Nevertheless, generalization still poses a challenge to the adoption of a solution based on neural networks. This has resulted in worse performance of the segmentation model, and even more so the survival model in the unseen cases of the test set.

The good generalization of Weibull AFT may due to the fewer parameters of this model compared with their deep learning counterparts, greatly reducing the possibility of overfitting. Some possible ways to mitigate this include reducing the neural network capacity (number of parameters), and to use \emph{n-fold} cross validation and regularization during training. On the other hand, the superior performance of Weibull AFT with respect to Cox PH can be attributed to the acceleration/deceleration effect of the input features on the hazard probability, rather than their time independence, as assumed by the Cox PH model.

Finally, we have incorporated tumour-instance information in the prediction via processing tumour descriptors and tumour centred image patches with the improved graph attention networks \cite{Brody2021}. Explanation algorithms like the approximated Shapley values \cite{Ancona2019} could be combined with the proposed method to increase the interpretability of predictions, a matter of crucial importance in clinical practice. Similar approaches have been used for gene expression data \cite{Hayakawa2022} and histopathology images \cite{Bhattacharjee2022}. We leave the application of these methods to head \& neck cancer as future work.

\end{document}